%% file: main.tex
\documentclass[5p,twocolumn]{elsarticle}

\usepackage{amsmath,amssymb,amsfonts}
\usepackage{algorithmic}
\usepackage{graphicx}
\usepackage{textcomp}
\usepackage{fontawesome}
\usepackage{makecell}
\usepackage{balance}

\newcommand{\totalLabeled}{681\xspace}

\newcommand{\conflicts}{344\xspace}
\newcommand{\totalLabels}{1,706\xspace}
\newcommand{\totalApps}{315\xspace}
\newcommand{\totalCommits}{4,781\xspace}

\newcommand{\participants}{43\xspace}

\newcommand{\textRevision}[1]{\textcolor{black}{#1}}

\input{preamble}

\begin{document}

\begin{frontmatter}
	
\markboth{Mazuera-Rozo et al.}{Taxonomy of Security Weaknesses in Java and Kotlin Android Apps}


\title{Taxonomy of Security Weaknesses in\\Java and Kotlin Android Apps}

\author{Alejandro Mazuera-Rozo$^{1,2}$, Camilo Escobar-Velásquez$^{2}$, Juan Espitia-Acero$^{2}$, David Vega-Guzmán$^{2}$,\\ Catia Trubiani$^{3}$, Mario Linares-Vásquez$^{2}$, Gabriele Bavota$^{1}$}

\address{$^{1}$Università della Svizzera italiana, Lugano, Switzerland, $^{2}$Universidad de los Andes, Bogotá, Colombia\\
$^{3}$Gran Sasso Science Institute, L'Aquila, Italy}

\begin{abstract}
	Android is nowadays the most popular operating system in the world, not only in the realm of mobile devices, but also when considering desktop and laptop computers. Such a popularity makes it an attractive target for security attacks, also due to the sensitive information often manipulated by mobile apps. The latter are going through a transition in which the Android ecosystem is moving from the usage of Java as the official language for developing apps, to the adoption of Kotlin as the first choice supported by Google. While previous studies have partially studied security weaknesses affecting Java Android apps, there is no comprehensive empirical investigation studying software security weaknesses affecting Android apps considering (and comparing) the two main languages used for their development, namely Java and Kotlin. We present an empirical study in which we: (i) manually analyze \totalLabeled commits including security weaknesses fixed by developers in Java and Kotlin apps, with the goal of defining a taxonomy highlighting the types of software security weaknesses affecting Java and Kotlin Android apps; (ii) survey \participants Android developers to validate and complement our taxonomy. Based on our findings, we propose a list of future actions that could be performed by researchers and practitioners to improve the security of Android apps.
\end{abstract}


\end{frontmatter}

\input{introduction}
\input{design}
\input{results}
\input{threats}
\input{related}

\input{conclusion}
\vspace{-0.4cm}
\section{Data Availability}
\vspace{-0.2cm}
The data used in our study are publicly available at \cite{replication}.
\vspace{-0.5cm}
\input{ack}


\balance

\bibliographystyle{elsarticle-num}
\bibliography{bibliography}

\end{document}

%% file: preamble.tex
\usepackage{amssymb}
\usepackage{hyperref}
\usepackage[plain]{fancyref}
\usepackage{ifdraft}

\usepackage[inline]{enumitem}
\usepackage[table]{xcolor}
\usepackage{xspace}
\usepackage[final]{listings}
\usepackage{acronym}
\usepackage{url}
\usepackage{amsmath}
\usepackage{amssymb}
\usepackage{booktabs} 
\usepackage{subfig}
\usepackage{balance}
\usepackage{dirtree}

\usepackage[ruled]{algorithm2e} 

\usepackage{etoolbox}
\makeatletter
\patchcmd{\@makecaption}
  {\scshape}
  {}
  {}
  {}
\patchcmd{\@makecaption}
  {\\}
  {.\ }
  {}
  {}
\makeatother

\definecolor{OliveGreen}{rgb}{0,0.6,0.3}

\renewcommand{\lstlistingname}{Snippet}
\newcommand*{\fancyreflstlabelprefix}{lst}
\newcommand*{\Freflstname}{\lstlistingname}
\newcommand*{\freflstname}{\MakeLowercase{\lstlistingname}}
\Frefformat{vario}{\fancyreflstlabelprefix}%
  {\Freflstname\fancyrefdefaultspacing#1#3}
\frefformat{vario}{\fancyreflstlabelprefix}%
  {\freflstname\fancyrefdefaultspacing#1#3}
\Frefformat{plain}{\fancyreflstlabelprefix}%
  {\Freflstname\fancyrefdefaultspacing#1}
\frefformat{plain}{\fancyreflstlabelprefix}%
  {\freflstname\fancyrefdefaultspacing#1}

\newcommand*{\fancyreflnlabelprefix}{ln}
\newcommand*{\Freflnname}{Line}
\newcommand*{\freflnname}{\MakeLowercase{\Freflnname}}
\Frefformat{vario}{\fancyreflnlabelprefix}%
  {\Freflnname\fancyrefdefaultspacing#1#3}
\frefformat{vario}{\fancyreflnlabelprefix}%
  {\freflnname\fancyrefdefaultspacing#1#3}
\Frefformat{plain}{\fancyreflnlabelprefix}%
  {\Freflnname\fancyrefdefaultspacing#1}
\frefformat{plain}{\fancyreflnlabelprefix}%
  {\freflnname\fancyrefdefaultspacing#1}

\lstdefinelanguage{JavaScript}{
keywords={typeof, new, true, false, catch, function, return, null, catch, switch, var, if, in, for, while, do, else, case, break, throw, this, instanceof},
keywordstyle=\color{purple}\bfseries,
ndkeywords={},
ndkeywordstyle=\color{blue}\bfseries,
identifierstyle=\color{black},
sensitive=false,
comment=[l]{//},
morecomment=[s]{/*}{*/},
commentstyle=\color{OliveGreen}\ttfamily,
stringstyle=\color{OliveGreen}\ttfamily,
morestring=[b]',
morestring=[b]"
}
\usepackage{color}
\definecolor{gray97}{gray}{.97}
\definecolor{gray90}{gray}{.90}
\definecolor{gray75}{gray}{.75}
\definecolor{gray45}{gray}{.45}
\definecolor{codegreen}{rgb}{0,0.6,0}
\definecolor{codered}{rgb}{0.6,0,0}
\definecolor{codegray}{rgb}{0.5,0.5,0.5}
\definecolor{codepurple}{rgb}{0.58,0,0.82}
\lstdefinestyle{floating}{%
  frame=none,
  float=htb,
  captionpos=b
}

\lstdefinestyle{ctxtraits}
 {language=JavaScript,
  frame=lines,
  showstringspaces=false,
  keywordstyle=\tt\bf,
  tabsize=3,
  style=floating,
  morekeywords={Trait, cop, Context, activate, deactivate, adapt, addObjectPolicy, manager}
}

\lstnewenvironment{ctxtraits}[1][]
 {\lstset{style=ctxtraits,#1}}{}

\newcommand{\ie}{\emph{i.e.,}\xspace}
\newcommand{\eg}{\emph{e.g.,}\xspace}
\newcommand{\etc}{etc.\xspace}
\newcommand{\etal}{\emph{et~al.}\xspace}
\newcommand{\secref}[1]{Section~\ref{#1}\xspace}

\newcommand{\figref}[1]{Fig.~\ref{#1}\xspace}

\newcommand{\tabref}[1]{Table~\ref{#1}\xspace}

\usepackage{ifthen}
\newboolean{showcomments}
\setboolean{showcomments}{true}
\ifthenelse{\boolean{showcomments}}
{\newcommand{\nb}[2]{
		\fbox{\bfseries\sffamily\scriptsize#1}
		{\sf\small$\blacktriangleright$\textit{#2}$\blacktriangleleft$}
	}
}
{\newcommand{\nb}[2]{}
}

\definecolor{author}{rgb}{.5, .5, .5}
\definecolor{comment}{rgb}{.1, .0, .9}
\definecolor{note}{rgb}{.9, .4, .0}
\definecolor{idea}{rgb}{.1, .7, .0}
\definecolor{missing}{rgb}{.9, .1, .0}
\definecolor{deleteme}{rgb}{.9, .1, .0}

%% file: introduction.tex

\section{Introduction} \label{sec:introduction}
\vspace{-0.1cm}

Mobile apps and devices are nowadays omnipresent in daily life activities, supporting many crucial  
		tasks (\eg banking, social networking, \etc) involving the manipulation and storage of sensitive and private data. The usage of mobile operating systems has already exceeded the usage of desktops/laptops operating systems \cite{statcounter2020a,statcounter2020b,google2019report}. As a consequence, mobile apps and devices have become a very attractive target for malicious attacks aimed at stealing private and sensitive information from apps/devices and to exploit on-device capabilities such as processing, data collection via sensors, and networking. Also, according to the CVE details portal \footnote{\url{https://www.cvedetails.com/product/19997/Google-Android.html}} the number of vulnerabilities in the Android operating system has seen a steep growth in the last years, with a total of 2563 reports in 10 years (2009-2019). As a natural reaction to such a rising of vulnerabilities in the mobile ecosystem, original equipment manufactures (OEMs), operating system designers (\eg Google), researchers, and companies have devoted efforts to improve the security of mobile OSs, devices and apps.

A paramount example is the volume of research focused on detecting vulnerabilities in Android apps (see \eg~\cite{arzt2014flowdroid,li2015iccta,sadeghi2017taxonomy,lee2017sealant,singleton2019firebugs,you2016reference, bello2019opia,ren2015hijacking,novak2015covertchannels,gadient2019securitysmells}). The Android OS and devices have been also investigated in the context of previous studies aimed at categorizing their security weaknesses and exploits~(\eg \cite{huang2015servershutdown, thomas2015securitymetrics, cao2015inputvalidation,wang2016systemserver, jimenez2016profiling, bagheri2018androidpermissions, meng2018survey, mazuera2019android}). Even datasets with malicious apps have been built~\cite{allix2016androzoo, Zhou2012Genome}.
%

Still, to the best of our knowledge, there is no comprehensive taxonomy of security weaknesses exhibited in Android apps. With security weaknesses we refer to flaws or gaps in a software that could be exploited to violate its security policy, thus eventually causing a disruption of the confidentiality, integrity, or availability of the system in question. \textRevision{As compared to desktop applications, Android apps may suffer of specific vulnerability types since they (i) run on a mobile device, thus usually collecting a larger amount of information about the user (e.g., location, video/audio, as well as biometric information); (ii) are built on top of a specific framework and programming model, that, as we will show, requires to carefully handle specific types of resources and components (e.g., Activities, Intents, Broadcast Receivers,  etc.); (iii)  despite  the Android OS is built on top of the Linux kernel, several modifications have been done to the kernel, and there is a set of specific OS layers built on top of the kernel that makes Android apps programming different from web and desktop app programming, even the programming model is different from the  iOS model. In this paper, we focus on Android apps written in Java and in Kotlin, the two main programming languages officially supported for the development of Android apps\footnote{https://developer.android.com/kotlin/first}.}

Despite previous individual efforts for analyzing, detecting and fixing specific sets of security weaknesses, the research community still lacks a body of knowledge characterizing the types of weaknesses affecting Android apps. Also, some of the empirical investigations performed in the past could become outdated due to the frenetic evolution of the Android ecosystem. Indeed, the programming models include now the possibility of creating native, hybrid, cross-platform, and mobile web apps for the Android platform. 
Previous studies on specific security vulnerabilities have focused on analyzing Android Java apps, because of the availability of code bases and APKs in this language. Given the rising interest for Kotlin apps and its status of official Android language, investigating security weaknesses in Kotlin becomes a required avenue for research. While Dart\footnote{\textRevision{Dart is a programming language developed by Google and designed to support the implementation of  applications, including mobile apps. https://dart.dev/}}/Flutter\footnote{\textRevision{Flutter is a software development kit created by Google that is built on top of Dart and can be used to develop cross-platform applications. https://flutter.dev/}} also represent interesting targets for research, their diffusion is still limited, with $\sim$18k GitHub repositories as compared to the $\sim$75k Kotlin repositories (May 2020).

In this paper, we present the first empirical study characterizing software security weaknesses in Android Java and Kotlin apps. To this end, we build a taxonomy of security weaknesses by (i) manually analyzing \totalLabeled commits in open source Android Java/Kotlin apps (\ie \emph{mining-based study}), and (ii) surveying \participants Android developers to collect their experience with security weaknesses, and in particular with the types they frequently faced (\ie \emph{survey-based study}). The output of the mining-based study is a taxonomy on multiple levels featuring a total of 74 categories of security weaknesses.

As results of the developers' survey, we identified 28 types of security weaknesses, of which 22 were already covered in our taxonomy, and six more were added. We use the defined taxonomy to discuss interesting directions for future research in the area, and lessons learned for practitioners.

Note that, while catalogues of security weaknesses in mobile apps have been previously defined \cite{cwemobile,OWASP}, they are not based on the empirical observation of weaknesses affecting real mobile apps and, as a result, they are less comprehensive than the taxonomy we derive in this work.

%% file: design.tex

\newcommand\rqone {What are the types of software security weaknesses faced by the developers of Java and Kotlin Android apps?}

\vspace{-0.2cm}
\section{Study Design} \label{sec:design}
\vspace{-0.2cm}
The {\em goal} of the study is to investigate software security weaknesses affecting Java and Kotlin Android apps. 
The {\em context} consists of (i) \totalLabeled commits performed by software developers of Android apps to fix software security weaknesses, and (ii) answers to a survey conducted with \participants Android developers to investigate the software security weaknesses they face and how they deal with their identification and fixing.

\noindent Our study addresses the following research question:
\vspace{-0.1cm}
\begin{quote}
\textbf{RQ$_1$:}\emph{ \rqone}\smallskip
\end{quote}
\vspace{-0.3cm}
To answer RQ$_1$, we combine two orthogonal analyses. We start by manually analyzing a set of \totalLabeled commits fixing security weaknesses performed in \totalApps Java and Kotlin open source Android apps with the goal of defining a taxonomy of software security weaknesses faced by Android developers. We analyze both apps written in Java and in Kotlin, by presenting the differences (if any) in the distribution of security issues across the two languages. Then, we run a survey with \participants Android developers. The survey has a dual goal. First, we ``validate'' the taxonomy defined in the first step, by asking developers which security weaknesses they address more often. This allows to assess the comprehensiveness of our taxonomy and to complement it with new categories of security weaknesses if needed. Second, we collect additional data reporting how developers perceive security weaknesses in Android apps.

\vspace{-0.2cm}
\subsection{Manual Analysis of Commits}
\label{sub:manualDesign}

We present the procedure to collect the data needed for our study (\ie commits fixing security weaknesses we manually validated) and the process performed to derive our taxonomy.

\vspace{-0.2cm}
\subsubsection{Data Collection}

As previously explained, Java has been historically the official programming language for creating Android apps. However, in 2019, Google announced that Kotlin is its official and preferred language for native Android apps.\footnote{\url{https://tcrn.ch/363AyBv}} Thus, when selecting the mobile apps to study, we made sure to have a mix of Java and Kotlin apps by (i) merging different datasets available in the literature, and (ii) mining  a dataset we created for this study. Keep in mind that for all considered apps we must have access to their repositories, since  we later mine their commits. Having in mind previously mentioned considerations, we adopted the three following datasets. \smallskip

\textbf{Geiger \etal\cite{pascarella2018osprojects}} This dataset is composed of 8,431 real-world open-source Android apps. It combines source and commit history information from GitHub with metadata from Google Play store. We processed the dataset to exclude apps that are no longer available on GitHub, leading to 7,862 apps currently usable from this dataset (all available both on GitHub and on the Google Play store). 

\textbf{Coppola \etal\cite{coppola2019migrationkotlin}} The authors of this dataset mined all projects hosted on F-Droid \footnote{\url{https://f-droid.org}}, a repository for free and open source Android apps. 

This dataset is interesting because Coppola \etal reported the presence of 19\% of apps featuring Kotlin code among the 1,232 mined apps. We excluded apps that are no longer available on GitHub and, for consistency with the previous dataset, also those not published in the Google Play store. This resulted in 472 projects.

\textbf{GitHub Archive.} Since in the two previous datasets there is a prevalence of Java apps (also due to the fact that they were built before the announcement by Google pushing Android apps towards Kotlin), we ran a query on GH Archive \footnote{\url{https://www.gharchive.org}} using Google BigQuery, with the goal of identifying repositories having Kotlin as the primary language. The query is available in our online appendix \cite{replication}. The aforementioned query was run on March 1st, 2020, obtaining a list of 3,967 repositories as a result. We sorted these projects by number of stars (in descending order) and selected the top 5\% (\ie 200 repositories) for manual analysis. In particular, we checked that the 200 repositories were real-world Android apps available on the Play Store. From this screening, we obtained a list of 22 Kotlin apps to consider in our dataset. \smallskip

We aggregated these three datasets and removed duplicates, obtaining a final list of 8,157 open-source Android apps. The list is available in our replication package \cite{replication}.
We cloned all 8,157 repositories and ran on them a customized version of {\tt git-vuln-finder} \cite{gitvulnfinder}, a Python application aimed at finding commits likely to fix a security weakness. The search is based on a set of regular expressions applied on the commit message \cite{zhou2017commits}. While most of the used regular expressions are applicable in the context of mobile apps, the work by Zhou and Sharma \cite{zhou2017commits} focuses on web applications. Thus, we modified their tool by complementing the list of regular expressions with others we defined by looking at the list of security weaknesses relevant to mobile apps and present in the Common Weakness Enumeration (CWE \footnote{\url{https://cwe.mitre.org}}) version 4.0, a community-developed list of common software and hardware security weaknesses. Also, we considered a commit as relevant for our study if it explicitly mentions the name or id of any weakness present in the CWE dictionary. The adopted regular expressions are publicly available \cite{replication}.

After running {\tt git-vuln-finder} on the 8,157 projects, we identified a set of candidate  commits from which we removed duplicates due to: (i) commits mined from both the master branch and other branches merged in the master; (ii) forked repositories. Also, we decided to keep in our dataset only commits in which the developers are modifying a single Java or Kotlin file (as identified by their extension). The rationale behind this decision is two-fold. First, if a developer mentions in the commit note that she is fixing a security weakness and only one file is modified in the commit, we can be sure that the fix happened in that file. Second, since we aim at classifying the type of security weakness involved in each commit, understanding a fix spanning across many files can be quite challenging, and lead to misclassifications. 

This cleaning process resulted in a final list of \totalCommits candidate commits.

\vspace{-0.2cm}
\subsubsection{Open Coding}

Given the \totalCommits commits collected in the previous step, we manually analyzed \totalLabeled of them with the goal of describing, using a label, the type of security weakness fixed in the commit. The number of inspected commits ensures a significance interval (margin of error) of $\pm5\%$ with a confidence level of 99\%. We did not use random sampling for the selection of the commits to manually inspect. Indeed, in the set of \totalCommits candidate  commits, there are 4,391 commits impacting a Java file, and 390 modifying a Kotlin file. Since we aim at comparing the types of security weaknesses affecting these two main languages used to develop native Android apps, we decided to target the analysis of the same number of Java- and Kotlin-related commits. We targeted the inclusion of 200 valid commits per language (\ie excluding commits labeled as false positive since they are not related to security weaknesses' fix). 

The choice of 200 was tailored on the amount of commits available for Kotlin, since we expected to find a substantial number of false positives as result of the regular expressions used to select the commits. By applying the process described in the following, we analyzed 360 Java-related commits (200 valid + 160 false positives) and 321 Kotlin-related commits (200 valid + 121 false positives).

Five authors took part to the labeling process that was supported by a web application. Each author independently labeled the commits randomly assigned to her/him by the web application, defining a ``label'' describing the security weakness fixed in each commit. To define such a label the authors manually inspected the diff of the commit and the message accompanying it. As a guideline for the label definition, the authors used the CWE 4.0 list. The authors reused as much as possible the list of security weaknesses in CWE, defining new labels only when needed.
%
Moreover, the web application also showed the list of labels created so far, allowing the author to select one of the already defined labels. Since the number of possible labels (\ie types of security weaknesses) is extremely high, such a choice helps using consistent naming while not introducing a substantial bias. In case the commit was not related to a security weakness fix, a \emph{false positive} label was assigned, discarding the commit from the study. Each commit was assigned to two authors and, in cases for which there was no agreement between the two authors, the commit was assigned to a third author. Conflicts arisen for \conflicts commits ($\sim$50\% of \totalLabeled). While such a number may look high, note that we considered as a conflict also cases in which the authors used two slightly different labels to express the same concept (\eg CWE-703: improper check or handling of exceptional conditions \emph{vs} CWE-754: improper check for unusual or exceptional conditions). A total of \totalLabels labels was required in order to reach our target of assessing and characterizing 200 valid commits per programming language: two labels per each of the 400 valid commits (800), two labels for each of the 281 false positives we discarded (562), and one more label for each of the \conflicts solved conflicts (\conflicts).

As outcome, we present a taxonomy of software security weaknesses identified in the manual analysis and we complement our discussion with qualitative examples. 

\begin{table}[ht]
\scriptsize
\centering
\caption{Structure of the survey used in our study\vspace{-0.3cm}}
\label{tab:survey}
\resizebox{\linewidth}{!}{
\rowcolors{2}{gray!15}{white}

\begin{tabular}{p{6.5cm}}

\toprule

\textbf{BACKGROUND QUESTIONS} \\\midrule
$Q_1$: In which country do you live?\\
$Q_2$: What is your current job position?\\
$Q_3$: How many years of programming experience do you have?\\
$Q_4$: How many years of programming experience do you have concerning native Android apps? Please specify overall/Java/Kotlin/Dart.\\
$Q_5$: How many years of programming experience do you have concerning the testing of native Android apps?\\\midrule

\textbf{EXPERIENCE WITH SOFTWARE security weaknesses  AND THEIR PERCEPTION} \\\midrule
$Q_6$: Which factors do you consider to estimate the likelihood of a security weakness to be exploited?\\
$Q_7$: Which factors do you consider to estimate the negative impact of a security weakness in case it is exploited?\\
$Q_8$: Which are the most common security weaknesses that you found?\\ 
$Q_9$: Which security weaknesses do you consider as the most dangerous?\\
$Q_{10}$: How do you detect security weaknesses? Do you use any specific tool for this task?\\\bottomrule

%
%

\end{tabular}
}
\end{table}

\vspace{-0.2cm}
\subsection{Survey with Developers}
\label{sub:surveyDesign}
We designed a survey aimed at investigating the types of security weaknesses that are found by developers in their apps and their perception about specific aspects of security weaknesses. The survey was designed to last at most 15 minutes, to maximize the survey completion rate. The survey structure is reported in \tabref{tab:survey}. Note that we rephrased some of the questions to shorten them. First, we collected background information about participants ($Q_1$-$Q_5$). If a participant answered ``zero'' to the part of $Q_4$ related to the overall programming experience of native Android apps \footnote{With \emph{native Android apps}, we refer to mobile apps written in one of the official programming languages of Android (\ie Java and Kotlin)}, the survey ended, and the participant was excluded from the study. This happened in 2 cases. 

Then, $Q_6$-$Q_7$ aimed to collect information about the developers' perception of security weaknesses. For these questions we provided a predefined list of possible factors to check, with the possibility of specifying additional factors. For $Q_6$, the predefined list included: \emph{Skill level required to exploit it}, \emph{Motivation to exploit it}, \emph{Chances for a successfully exploit}, \emph{Number of agents needed for the exploit\footnote{\textRevision{With “agents needed for the exploit” we refer to the number of attackers that are needed to exploit a security weakness. Indeed, not all security issues can be exposed by a single attacker}}}, \emph{Ease of discovery}, \emph{Technical difficulty of the exploit}, \emph{How well-known is the weakness}, and \emph{How likely is the exploit to be detected}. Concerning $Q_7$: \emph{Confidentiality}, \emph{Integrity}, \emph{Availability}, \emph{Accountability}, \emph{Brand reputation}, \emph{Business profits}, and \emph{Privacy violation}. 

$Q_8$ and $Q_9$ aimed to validate/complement the taxonomy defined as output of the manual study, with $Q_8$ focusing on the most frequent and $Q_9$ on the most dangerous security weaknesses experienced by developers. Both these questions required an open answer. Two authors read each answer and assigned the CWE-ID(s) needed to describe the security weaknesses mentioned in each answer. A third author merged these tags and solved conflicts arisen for 15 answers (18\%). 

Since a respondent might have answered the same for $Q_8$ and $Q_9$, duplicates among these answers were removed to avoid counting twice the same security weakness mentioned by the same developer. 

Finally, $Q_{10}$ asked developers how they detect security weaknesses and whether they are supported by any tool.

\begin{figure}[ht]
	\vspace{-0.4cm}
	\centering 
	\hspace{-0.7cm}\includegraphics[width=1.07\linewidth]{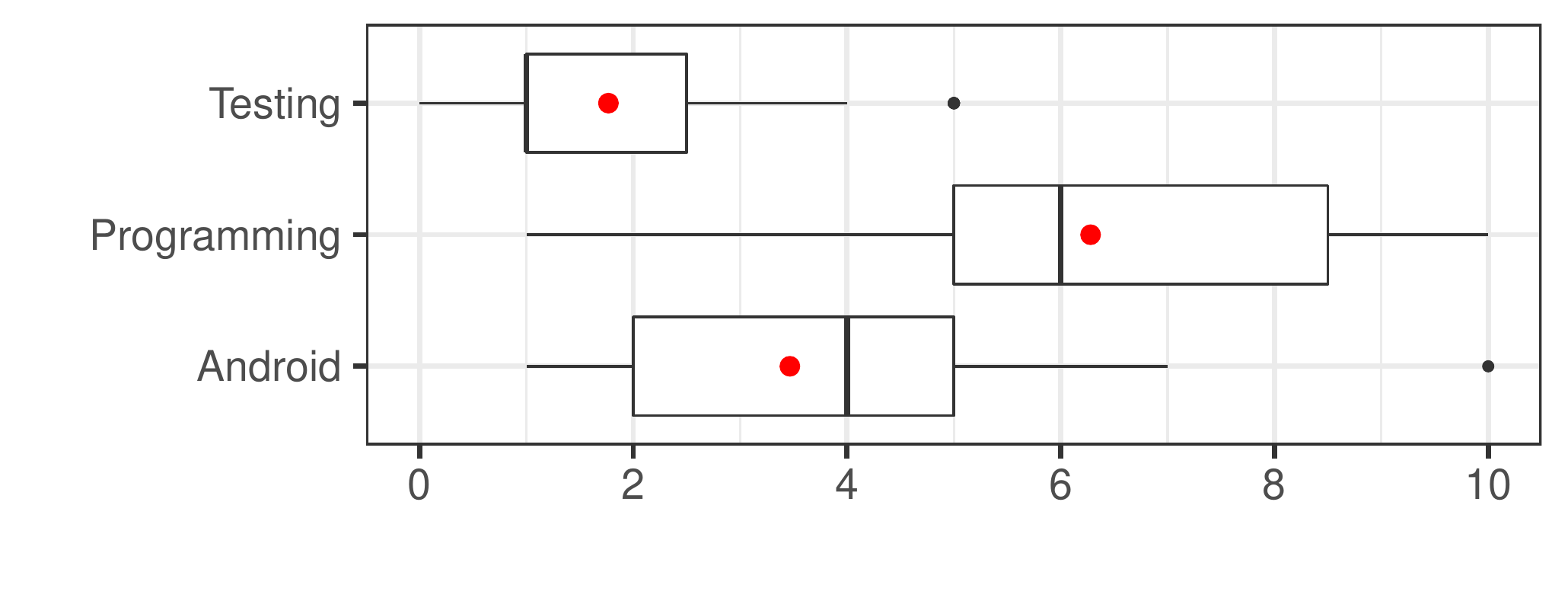}
	\vspace{-0.9cm}
	\caption{Experience in years of the \participants surveyed participants.}
	\label{fig:demographics}
\vspace{-0.7cm}
\end{figure}

We used convenience sampling to invite developers from companies we know to participate in our survey. Also, the link to the survey was shared in social media.
We collected answers for ten days, with a total of \participants participants that completed our survey from nine countries (\ie Argentina, Canada, Colombia, Germany, Hungary, Italy, Macedonia, Poland and USA). On average, the participants had $\sim$6 years of overall programming experience and approximately 3 years of Android development experience (see \figref{fig:demographics}). The average testing experience is close to two years. Regarding their job position, 21\% of participants are B.Sc. students, 7\% M.Sc. students, 4.6\% Ph.D students and 67.4\% professional Android developers having different job positions in the industry (\eg Senior Android developer, Technical leader, Project Management Engineer, Director).

\noindent

\subsection{Testing the Generalizability of Our Taxonomy}

Once obtained the final taxonomy including both categories defined through the mining-based study as well as those complemented by the developers' survey, we assessed its generalizability. We used all 64 Kotlin-related commits we did not manually analyze while building our taxonomy and a sample of 186 Java-related (again, among those we did not analyze). Then, we asked two Master students both having experience in Android development and not involved in the taxonomy definition and unaware of its structure, to perform the same manual analysis previously described. Each of them independently evaluated all instances. Conflicts arisen in 68\% of cases were solved through an open discussion between the two students and the first two authors of this work. The final output is a taxonomy of security weaknesses affecting Android apps, that we can compare with the taxonomy we defined to assess its stability. While in principle more Kotlin-related commits would be needed, we labeled  all those we found by mining several datasets of Android apps. 

\vspace{-0.2cm}
\subsection{Data Analysis}
\label{sub:dataAnalysis}

We start by presenting the taxonomy of types of software security weaknesses output of our mining-based study. Then, we discuss how the developers' survey helped in validating/complementing the obtained taxonomy. Finally, we report about the results of the generalizability study. The data used in our study are publicly available \cite{replication}. 

%% file: results.tex

\begin{figure*}
\begin{center}
\includegraphics[width=\linewidth, angle =90, scale=1.25]{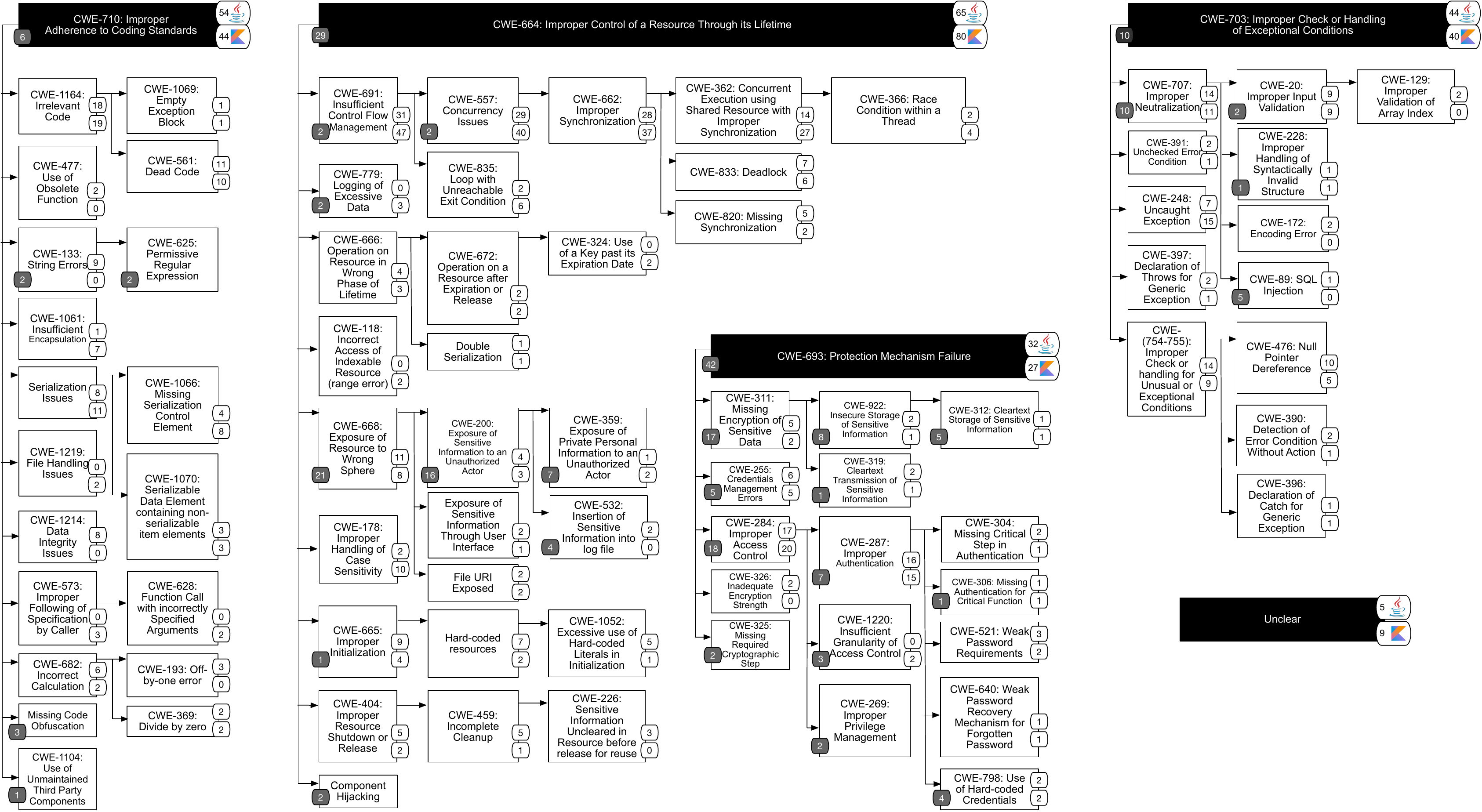}
\caption{Types of security weaknesses found in Java and Kotlin Android apps.}
\label{fig:taxonomy}
\end{center}
\end{figure*}

\section{Results} \label{sec:results}
\vspace{-0.2cm}
\figref{fig:taxonomy} depicts the taxonomy presenting the types of security weaknesses we found. Each sub-hierarchy uses a different color, with the black boxes representing the root categories. 

The taxonomy is derived by a total of 400 commits (200 for Java and 200 for Kotlin) we manually validated. However, there are 14 commits that were grouped in the \emph{Unclear} category since in these cases, while it was clear the intent of fixing a security flaw, we were unable to derive the  type of fixed security weakness. Each category in \figref{fig:taxonomy} is accompanied by one, two, or three numbers. The two numbers with white background represent the number of instances of the corresponding security weakness type we found in Java (top number) and Kotlin (bottom). The one with gray background represents  the number of developers that mentioned the type of security weakness in our survey. Categories added to our taxonomy as the result of the survey (\eg \emph{CWE-625: Permissive Regular Expression}), only have a gray-background number. 

\textRevision{It is worth noting that some categories have only been found in a few commits or have only been mentioned by developers (but not found in the mining-based study). Concerning the first case (\ie low number of commits related to the category), we preferred to still include those categories since, thanks to the numbers attached to them, it is easy for the reader to assess their relevance. In other words, it is clear from our taxonomy that, for example, the prevalence of \emph{CWE-691} vulnerabilities (78 overall instances) is much higher as compared to \emph{CWE-779} (3 overall instances). Concerning the latter case (\ie categories only mentioned by developers), they increase the comprehensiveness of our taxonomy; the fact that we did not find them in the analyzed sample of commits does not make them less relevant for our study. Indeed, while we analyzed a substantial set of commits (400), it is reasonable to expect that we did not encounter specific types of vulnerabilities in our study (as we will also show in \secref{sec:stability}).}

In addition, it is worth mentioning the hierarchical organization of the categories, moving from the most general categories (\ie the root nodes, such as \emph{CWE-710}), to more specialized ones (\eg \emph{CWE-1164}) down to the leafs (\eg \emph{CWE-1069}). The sum of instances for all child categories of a given node is lower or equal than the number of instances reported in its parent node. For example, \emph{CWE-1069} and \emph{CWE-561} are the two child categories of \emph{CWE-1164} (top-left corner of \figref{fig:taxonomy}). \emph{CWE-1069} and \emph{CWE-561} have 1 and 11 Java-related instances, respectively, while their parent category \emph{CWE-1164} has 18 Java-related instances. This is due to the labeling process since we assigned to each commit the most specific security weakness type we could derive from the manual inspection. Thus, for 12 of the 18 commits belonging to \emph{CWE-1164} we managed to provide a more specific categorization, resulting in the two child categories, while for 6 of them \emph{CWE-1164} was the most detailed label we could assign. Finally, some categories are not linked to any CWE-ID. These categories are either (i) aggregating some sub-categories for better visualization, or (ii) created by the authors since they did not find a proper type within the CWE dictionary to classify an instance (See  \tabref{tab:newVulns}).

\begin{table*}[tb]
	\centering
	\caption{Definition of categories created by authors not existing in CWE dictionary}
	\label{tab:newVulns}
	\rowcolors{2}{gray!15}{white}
	\begin{tabular}{p{4.9cm}|p{12.3cm}}
		
		\toprule
		
		\textbf{Categories} & \textbf{Common consequences } \\ \midrule
		
		Missing Code Obfuscation &\textRevision{ Non-obfuscated code is susceptible to reverse engineering allowing an attacker to retrieve sensitive information from a system.} \\ 
    	Double Serialization & \textRevision{Data mishandling can lead to a degradation of its integrity quality.}\\
		Exposure of Sensitive Information Through User Interface & \textRevision{Sensitive information could be exposed within the GUI to an actor that is not explicitly authorized to have access to that information.} \\
		File URI Exposed & \textRevision{A file can be made unsafely accessible from other apps providing unintended actors with inappropriate access to the resource.} \\
		Component Hijacking & \textRevision{A vulnerable component within an app can be seized by an actor to gain privileges in order to conduct operations originally prohibited.} \\
		
		\bottomrule
		
	\end{tabular}
	\vspace{-0.3cm}
\end{table*}

We start by discussing the categories output of the manual analysis (\secref{sec:mining}), presenting then the main differences between Java and Kotlin-related security weaknesses (\secref{sec:comparison}), and then discussing how the developers survey validated/complemented our taxonomy (\secref{sec:survey}). Finally, we present the results of the further manual validation performed by two Master students to test the generalizability of our taxonomy. We use icons to highlight parts related to implications for researchers (\faFlask) and practitioners (\faCodeFork).

\subsection{Mining-based Study} \label{sec:mining}

\emph{1) Improper Control of a Resource Through its Lifetime (145 instances - 36.25\%).}
It includes security weaknesses related to not maintaining or incorrectly maintaining control over a resource throughout its lifetime of creation, use, and release, leading to potentially exploitable states. 

A strongly represented type in this category is \emph{CWE-557: Concurrency Issues}, being prominent in both Java (29 instances) an Kotlin (40). 

\begin{figure}[ht]
\vspace{-0.2cm}
\begin{center}
		\includegraphics[width=1\linewidth]{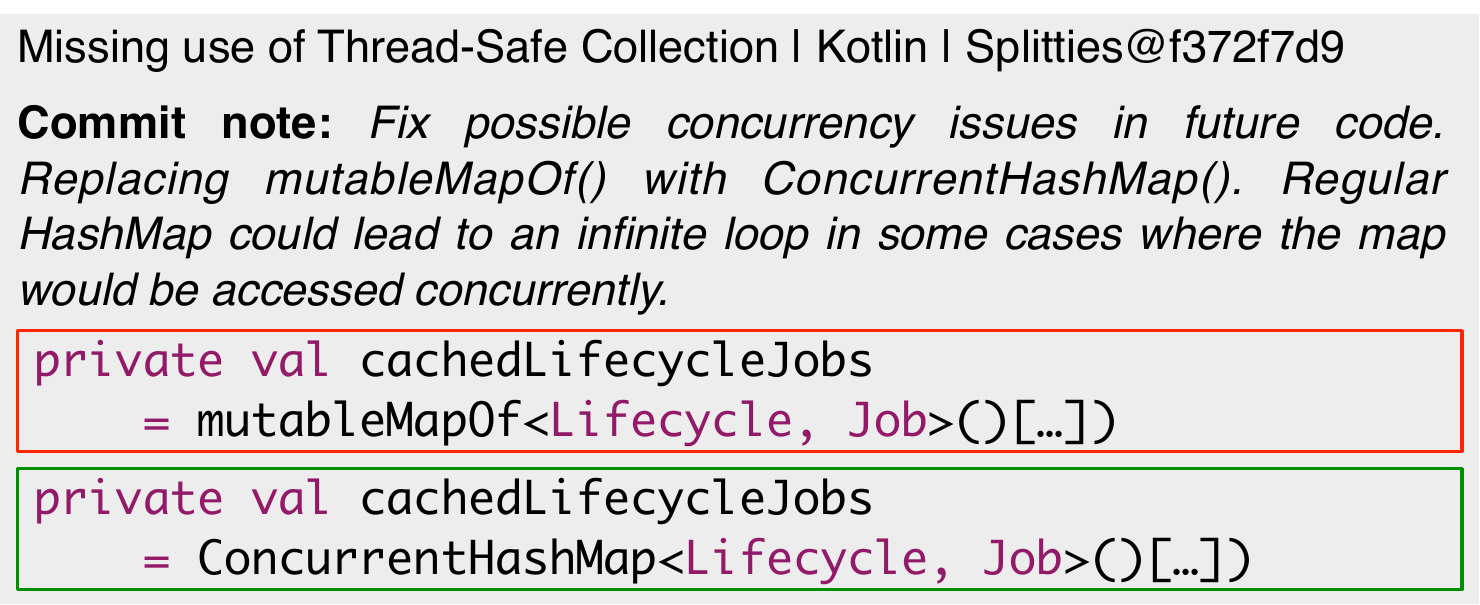}
		\caption{Usage of Thread-Safe Collection.}
		\label{fig:concurrency-557-3571}
\end{center}
\vspace{-0.3cm}
\end{figure}

\figref{fig:concurrency-557-3571} depicts an example of a concurrency issue in Kotlin code in which the developer is modifying the nature of the collection being used. 

The collection type {\tt mutable\-Map\-Of} is replaced with a {\tt Concurrent\-Hash\-Map}, preventing concurrency issues. The automatic detection and fixing of the type of code issues reported in the example can be easily targeted through approaches supporting \emph{Change Variable Type} refactoring. 

\faFlask~A customization of these techniques is needed to embed a set of ``change type'' rules that are relevant for security weaknesses (\eg replace {\tt mutable\-Map\-Of} with {\tt Concurrent\-Hash\-Map} if the class extends {\tt Thread}). 

\begin{figure}[ht]
\vspace{-0.2cm}
\begin{center}
		\includegraphics[width=1\linewidth]{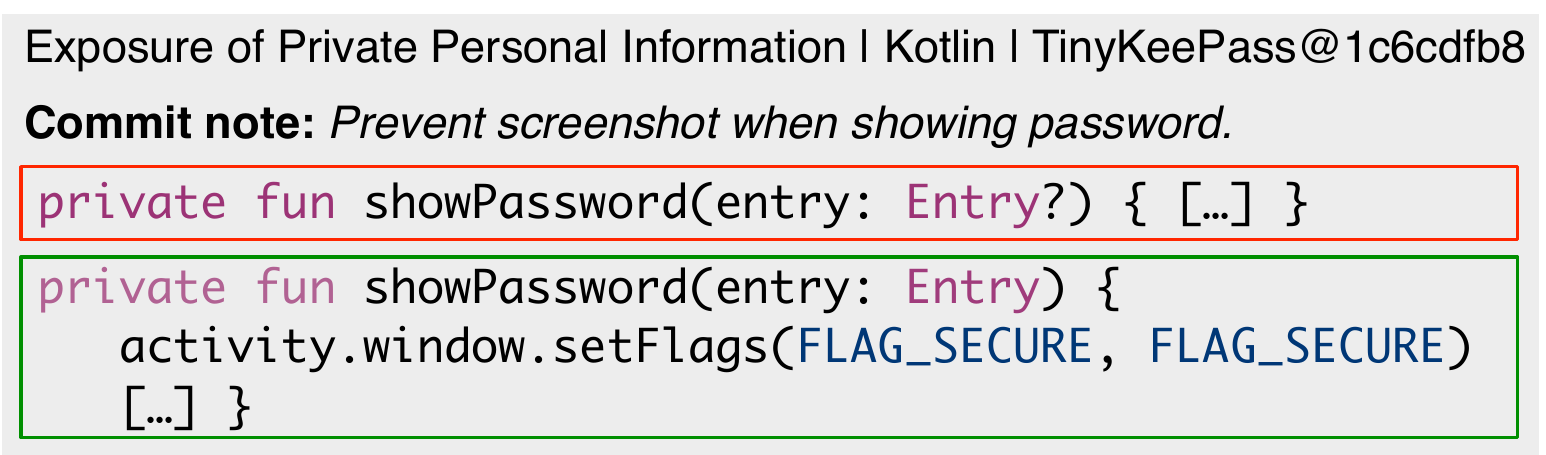}
		\caption{Exposing private information in the interface.}
		\label{fig:exposure-359-6381}
\end{center}
\vspace{-0.3cm}
\end{figure}

Another common weakness related to the improper control of resources is \emph{CWE-668: Exposure of Resource to Wrong Sphere}, with 11 instances found in Java and 8 in Kotlin. CWE-668 arises when a resource is inadvertently exposed due to insecure permissions or unexpected execution scenarios. \figref{fig:exposure-359-6381} shows Kotlin code in which the developer sets the {\tt FLAG\_SECURE} to a window showing a password in the app. 


The added flag asks the window manager to disable screen recording/capturing when the {\tt show\-Password} method is executed. The usage of this flag in windows containing sensitive information is recommended in the official Android documentation. Also in this case, \faFlask~techniques can be developed by researchers to automatically identify features in code that (i) deal with sensitive information that can be detected through simple keyword matching mechanisms (\eg looking for words like ``password''), and (ii) are in charge of displaying windows. Then, a {simple} automatic addition of proper flags can avoid potential points of attack. Such a security issue is also documented in Stack Overflow \footnote{\url{https://stackoverflow.com/questions/9822076}}. 

\faCodeFork~This suggests the potential usefulness for developers of recommender systems able to point out them to relevant Stack Overflow discussions while writing code (\eg Prompter\cite{Ponzanelli:emse2016}). Making the developer aware of such issues at coding time can avoid the introduction of the security flaw in the first place. 

Other types of security weaknesses that are less diffused but still relevant in the context of controlling resources are: \emph{CWE-178: Improper Handling of Case Sensitivity} (12 cases) and \emph{CWE-665: Improper Initialization} (13). The complete dataset of labeled weaknesses is available in the replication package \cite{replication}.

\emph{2) Improper Adherence to Coding Standards (98 instances - 24.50\%).} This category frames security weaknesses present in software due to ignored development best practices. The most represented sub-category for both programming languages is \emph{CWE-1164: Irrelevant Code}, with 18 Java and 19 Kotlin instances. This category is related, for example, to the presence of dead code in the apps (\ie code that is not executed in any of the app's feature). Such a code, while not executed in the normal app's usage, can still be unintentionally invoked/tested by software developers, or even exploited and executed by an attacker. The execution of dead code can be particularly dangerous since it is often not maintained with the latest security-related updates. Moreover, notice that for both investigated languages, dead code is not removed from the APK (\ie the compiled app) after compilation. \textRevision{Besides being possibly exploited, dead code can ``come back to life’’ by mistake, thus leading to unexpected consequences. For example, the implementation of a new feature can by mistake be invoking an old (dead) implementation of a method accessing the database, leading to a loss of information when the app is deployed. In addition, dead code ``might indirectly make it easier to introduce security-relevant weaknesses or make them more difficult to detect." \footnote{https://cwe.mitre.org/data/definitions/561.html}}. When dead code is identified, two strategies are usually adopted by developers to remove it \cite{Romano:tse2020}: (i) adding a explanatory comment before the dead fragment in question in which the developer mentions that the fragment it is or could be dead; and (ii) commenting out the code, leaving it available for future usage.  The latter strategy is the one that has been applied in one of the fixing commits we inspected. The developer is commenting out dead code that seems to be related to the management of contacts in the database. Two days before this commit, the same developer added a comment on top of the dead code saying {\tt //TODO: what is this for again?} (see changes to file {\tt MVP\_\-Activity\_\-Contacts} in commit {\tt f0801d88}). 

%

\faFlask~The prevalence of \emph{CWE-561: Dead Code} weaknesses in our taxonomy confirms the importance for researchers to investigate approaches able to automatically identify code components that can be removed without ripple effects on the code functionalities. 

\faCodeFork~To the best of our knowledge, very few tools are available for this task such as the one by Romano \etal~\cite{Romano:tse2020},  the Android Lint tool~\cite{ALINT},  and the Kotlin DCE plugin~\cite{DCE}.



Another prevalent type of security weaknesses related to coding standards are the ones grouped in the \emph{Serialization issues} category. \faCodeFork~A simple yet frequent issue we identified is the lack of a unique {\tt serial\-Version\-UID} in serializable classes, something expected in Java. Indeed, this identifier is stored with the serialized object and it is verified when deserializing it, thus to avoid data integrity issues.


All other first-level subcategories in the ``coding standard'' tree have less than ten total instances and, in several cases, are only related to one of the investigated languages (see \figref{fig:taxonomy}). 

The categories added as result of our survey will be discussed in \secref{sec:survey}.

\vspace{0.1cm}
\emph{3) Improper Check or Handling of Exceptional Conditions (84 instances - 21\%).}
This category includes weaknesses that can lead to unpredictable behavior due to the improper or missing handling of exceptional conditions rarely occurring during the normal operation of the app. Within this category, the most represented type of security weakness is \emph{CWE-707: Improper Neutralization}, happening when messages and/or data are not properly checked to be well-formed, valid, or benign (\ie the exceptional condition of malformed messages/data is not properly handled). This category is mostly composed by cases related to \emph{CWE-20: Improper Input Validation} (\eg issues related to the improper validation of the password in a login form, such as commit {\tt 4875515b} in the ccomeaux/boardgamegeek4android app, which could lead to a future credential management error). This type of issues can be addressed by relying on dynamic analysis, and in particular on fuzz testing, which aims at feeding unexpected input data that may generate crashes, exploit security weaknesses, or induce unexpected states in the~app. 

\faCodeFork~Several tools for this scope exist nowadays \cite{arzt2014flowdroid, monkey, DroidFuzzer, Huang2019fuzzing, EVOTAINT, IVDROID, DifFuzz}, thus giving to practitioners a vast repertory of available options that can be adopted for their testing activities. \textRevision{However, \faFlask~these tools work on Java and to the best of our knowledge, there are neither proposals of fuzzers that work at source-code level for Kotlin nor Dart/Flutter. In the case of Kotlin, fuzzers at the Java bytecode level could be used, however, this is not the case for Dart/Flutter apps because the  Dart language is not JVM-based.} Therefore, we encourage the research community to devise fuzzers and benchmarks for Kotlin and Dart such as \eg FuzzBench~\cite{FuzzBench}.

Another well-represented subcategory is \emph{CWE-248: Uncaught Exception}, that may cause the program to crash and/or expose sensitive information. Uncaught exceptions are a well-known issue in Android apps, especially when apps strongly rely on Android abstractions (\eg activities, asynctasks, \etc) \cite{Oliveira:jss2018}. The prevalence of this type of weakness in our taxonomy, \faFlask~supports previous findings reported in the literature, \faCodeFork~and highlights the potential usefulness for developers of tools developed in academia to automatically test Android apps using systematic input generation (see \eg \cite{linan2018rip,li2017droidbot}).

%
%

\vspace{0.1cm}
\emph{4) Protection Mechanism Failure (59 instances - 14.75\%).}
These security weaknesses are related to the incorrect restriction of access to a resource from an unauthorized actor. 

\begin{figure}[ht]
	\vspace{-0.3cm}
	\centering
	\includegraphics[width=0.75\linewidth]{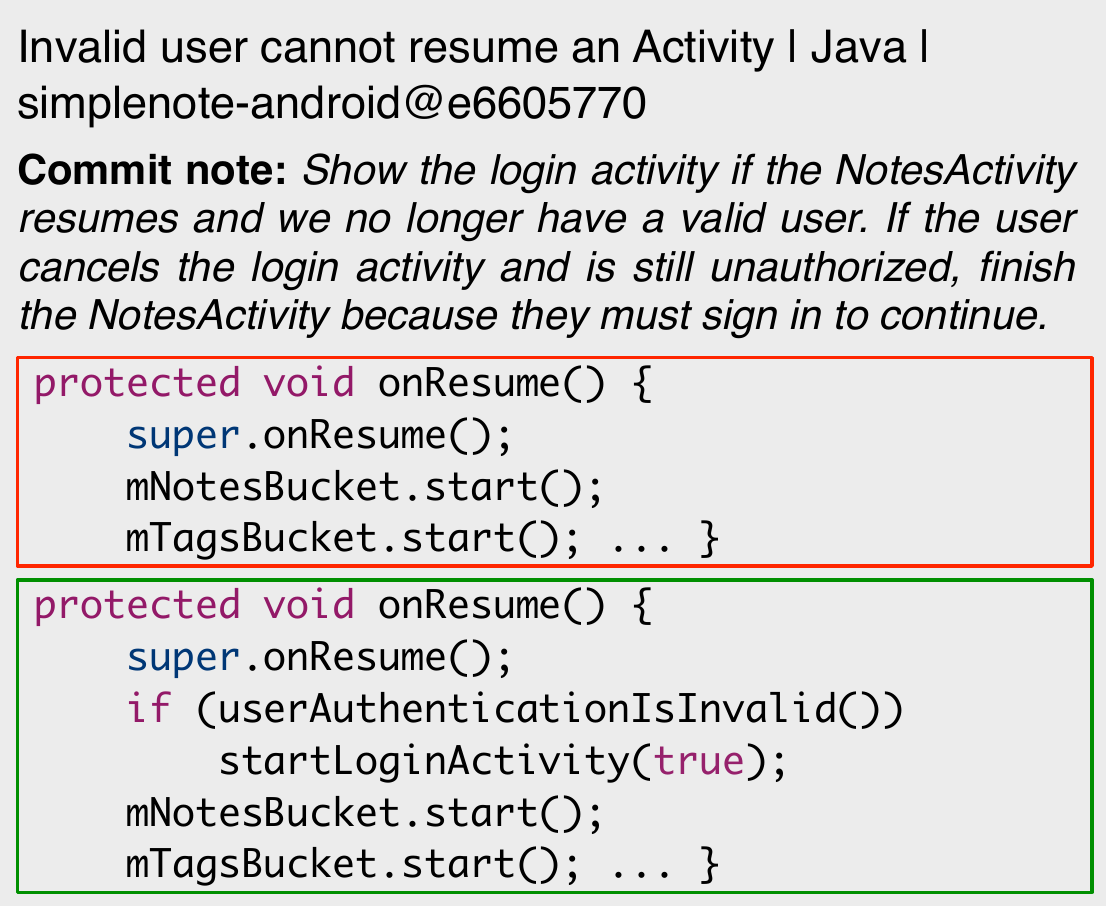}
	\caption{Unauthorized user must not be able to resume an Activity.}
	\label{fig:impr-auth-287}
\end{figure}

Thus, an attacker can compromise the security of the app by gaining privileges, accessing sensitive information, \etc Most of the weaknesses in this category are related to \emph{CWE-287: Improper Authentication}.  \figref{fig:impr-auth-287} shows an example of this type of security weakness, in which the developer fixes a security bug due to the missing authentication step in a feature requiring the user to have a valid authorization. \newline


While most of the cases in this category are simple programming mistakes (\eg a wrong/missing {\tt if} statement), these bugs are difficult to catch and automated testing tools are of little help here, since mostly focused on identifying app crashes. \faFlask~The development of approaches relying on machine learning (ML) to automatically discriminate apps' features that can be accessed with/without authentication could help. 
\newpage
This would require the existence of a large training set of apps components (\eg GUIs) labeled with their need for authentication (\eg a simple boolean). Assuming the feasibility of this ML-based approach, exploratory testing can then be used in combination with it to identify apps' features that should not be accessible without authentication but that, instead, can be reached by randomly exploring the app without a valid authentication. 



In this category we also found cases related to \emph{CWE-798: Use of Hard-coded Credentials}, such as commit {\tt f92221f} from the UserLAnd app. \footnote{\url{https://github.com/CypherpunkArmory/UserLAnd/commit/f92221f}} \faCodeFork~These cases are mostly due to hard-coded credentials most likely for testing purposes. However, using these credentials and having them available in repositories and/or URLs could lead to attacks.


\begin{figure}[ht]
	\begin{center}
		\includegraphics[width=0.8\linewidth]{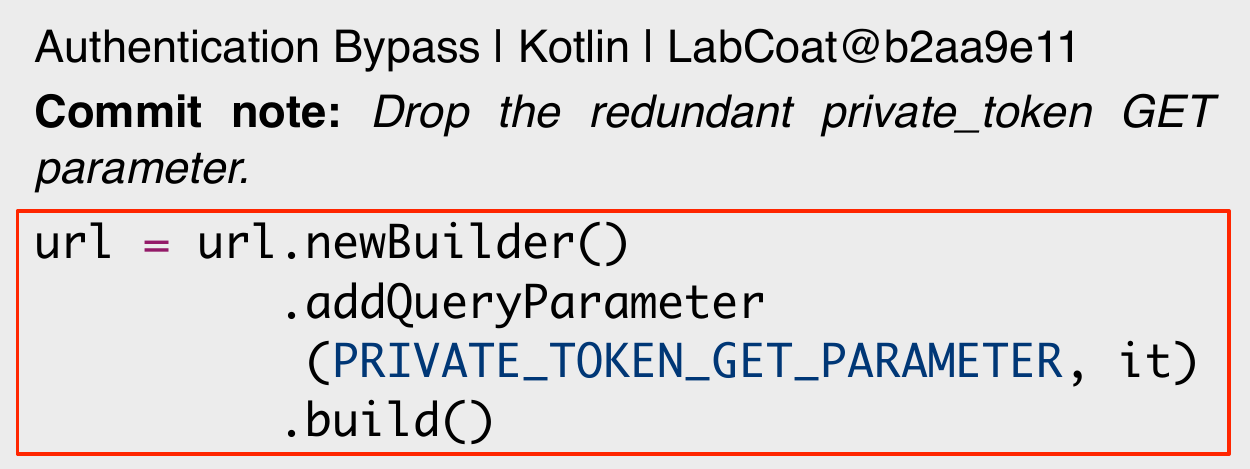}
		\caption{Exposing token within GET request.}
		\label{fig:capture-and-replay-294-3770}
	\end{center}
\vspace{-0.3cm}
\end{figure}

Finally, representative of the \emph{Improper Access Control} category is also the commit in \figref{fig:capture-and-replay-294-3770} . \textRevision{Before the fix, the app was sending a private token as a parameter within a GET request, which makes the token visible to anybody that can catch the URL, then exposing it and potentially allowing a Man-in-the-Middle attack, therefore under such circumstances an attacker could impersonate the user to whom the token belongs.} The commit fixes this issue by removing the problematic code.

In the case of the  example, a private token was sent as a parameter of a GET request, which makes the token visible to anybody that can catch the URL. This type of tokens are user to make the server know that the client sending  the  HTTP message is a valid/certified/authorized client, in that sense if  authentication tokens are visible when sending  an HTTP message, the user sending the token can be impersonated.

The identification of leaks for security-related information in mobile apps is an active research area\cite{bello2019opia}, with approaches extracting data from the apps' local databases and shared preferences to identify sensitive information that is not properly encrypted and/or anonymized. 

\faFlask~Identifying security-related information passed through query strings in URLs is a needed complement to these approaches.

\vspace{-0.2cm}
\subsection{Java vs Kotlin.} \label{sec:comparison}
This section compares the distribution of security weaknesses we observed in Java and Kotlin code. We focus on second-level categories (\ie the direct child nodes of the root categories). We do not consider in this discussion categories in which there are less than ten overall instances when summing up the weaknesses for Java and Kotlin. Indeed, whatever observation made for these categories may be due to the low number of instances in the category. Also, it is worth noting that our goal is simply to highlight the differences we found in our taxonomy. Indeed, explaining the reasons for the observed differences without best-guessing is not possible with the available empirical data. A different experimental design targeting this RQ is needed to properly answer it.

We found a balanced distribution of Kotlin/Java instances among most of the subcategories. In particular, no major differences are observed in the subtree related to \emph{CWE-710: Improper Adherence to Coding Standards}. 
Instead, when moving to the \emph{CWE-664: Improper Control of a Resource Through its Lifetime} subtree, we observe a slight prevalence of Kotlin-related security weaknesses. 

This is mostly due to more issues related to improper thread synchronization and handling of case sensitivity (\ie the code does not properly handle differences in case sensitivity, possibly leading to inconsistent results). 

Concerning the \emph{CWE-703: Improper Check or Handling of Exceptional Conditions} tree, the main category exhibiting differences is the one related to uncaught exceptions, with a prevalence of Java-related security weaknesses (15~\emph{vs}~7).

Finally, no major differences have been observed for what concerns \emph{CWE-693: Protection Mechanism Failure}.

Summarizing, the distribution of types of security weaknesses across Java and Kotlin seems to be quite similar. \faFlask~This suggests that previous findings reported in empirical studies about security weaknesses in Java Android apps are likely to generalize to Kotlin apps as well, at least for what concerns the security weaknesses diffusion.

\subsection{Survey with Developers} \label{sec:survey}

Our taxonomy has been validated/complemented through the survey we performed with software developers. In the developers' answers to $Q_8$ and $Q_9$ (see \tabref{tab:survey}), we found mentions to 87 software security weaknesses, that can be classified into the 28 types labeled with a gray number (\ie the number of developers who mentioned that security weakness type) in \figref{fig:taxonomy}. Out of these, 22 were already part of our taxonomy as output of the \emph{mining-based} study, while six were added: \emph{CWE-269: Improper Privilege Management}, \emph{CWE-325: Missing Required Cryptographic Step}, \emph{CWE-625: Permissive Regular Expression}, \emph{CWE-1104: Use of Unmaintained Third Party Components}, \emph{Hijacking}, and \emph{Missing Code Obfuscation}. The fact that 78\% of the security weakness types mentioned by developers (22/28) were already part of our taxonomy, provides a good level of confidence about its comprehensiveness. 

The most common security weaknesses ($Q_8$) mentioned by the surveyed developers can be easily seen in \figref{fig:taxonomy}, with those belonging to the \emph{CWE-693: Protection Mechanism Failure} and \emph{CWE-664: Improper Control of a Resource Through its Lifetime} trees representing 81\% of the mentioned security weaknesses (71/87). There is a common thread we found when analyzing the answers provided to $Q_9$, meaning the most dangerous weaknesses perceived by developers. All developers are mostly worried about unauthorized access to sensitive, private data stored in the app or sent/received through/by it. Some of the (shortened) answers: ``\emph{vulnerabilities related to confidentiality, since they can expose user information}'', ``\emph{wrong/missing encryption of data being stored within the app}'', ``\emph{the leak of user personal information}''. 

\faFlask~Answers to $Q_9$ confirm the importance of research studying security weaknesses related to data stored/manipulated by the apps~\cite{arzt2014flowdroid,bello2019opia,Blackdroid}.

An orthogonal view about the harmfulness of security weaknesses as perceived by developers is given by the answers to $Q_6$ (\ie the factors impacting the likelihood of a security weakness to be exploited) and $Q_7$ (\ie the factors impacting the harmfulness of the security weakness if exploited). 

Developers pointed to technical aspects when answering $Q_6$, indicating the difficulty of exploiting a security weakness as more important than the motivation to exploit  it (\ie the actual gain an attacker gets). Indeed, the difficulty of exploiting has been mentioned by 79\% of the surveyed developers, as compared to the $\sim$56\% mentioning the potential gain. Answers to $Q_7$ stress again the importance for developers of protecting sensitive information, with most (88.3\%) of the respondents reporting confidentiality and privacy violations as the main factors impacting the dangerousness of a security weakness. 

Finally, we analyze the answers provided for $Q_{10}$, related to the tools used by developers to detect security weaknesses. None of the surveyed developers mentioned tools developed in academia. Clearly, this does not mean that the adopted tools do not use any idea proposed in the literature.

Among the mentioned ones (available in our replication package) there are AppScan from IBM \cite{appScan}, Infer from Facebook \cite{infer}, Sonarqube \cite{sonarqube}, and pre-launch reports given by Google Play when uploading the app to the market. Then, we looked into the relevant literature for tools that can be used by developers to detect the types of security weaknesses they more often face or they perceive as more dangerous (\ie previously analyzed answers to $Q_8$ and $Q_9$). \tabref{tab:toolsSurveyess} reports categories of security weaknesses with corresponding references presenting approaches for their detection. Some categories are merged in a single row since their security weaknesses are quite similar, and approaches related for one category should work for the other as well. \faCodeFork~For 12 of the 28 types of security weaknesses mentioned by developers we found at least one approach supporting their automatic detection. \faFlask~On the one side, this shows that the research community is working on security weaknesses that are relevant for developers. 
On the other side, the developed approaches are unknown (at least) to our small pool of surveyed developers. This may also be due to the unavailability of industry-strength products implementing these approaches.

\subsection{Stability of the Taxonomy} \label{sec:stability}

\begin{figure*}
	\begin{center}
		\includegraphics[width=\linewidth, angle =90, scale=1.25]{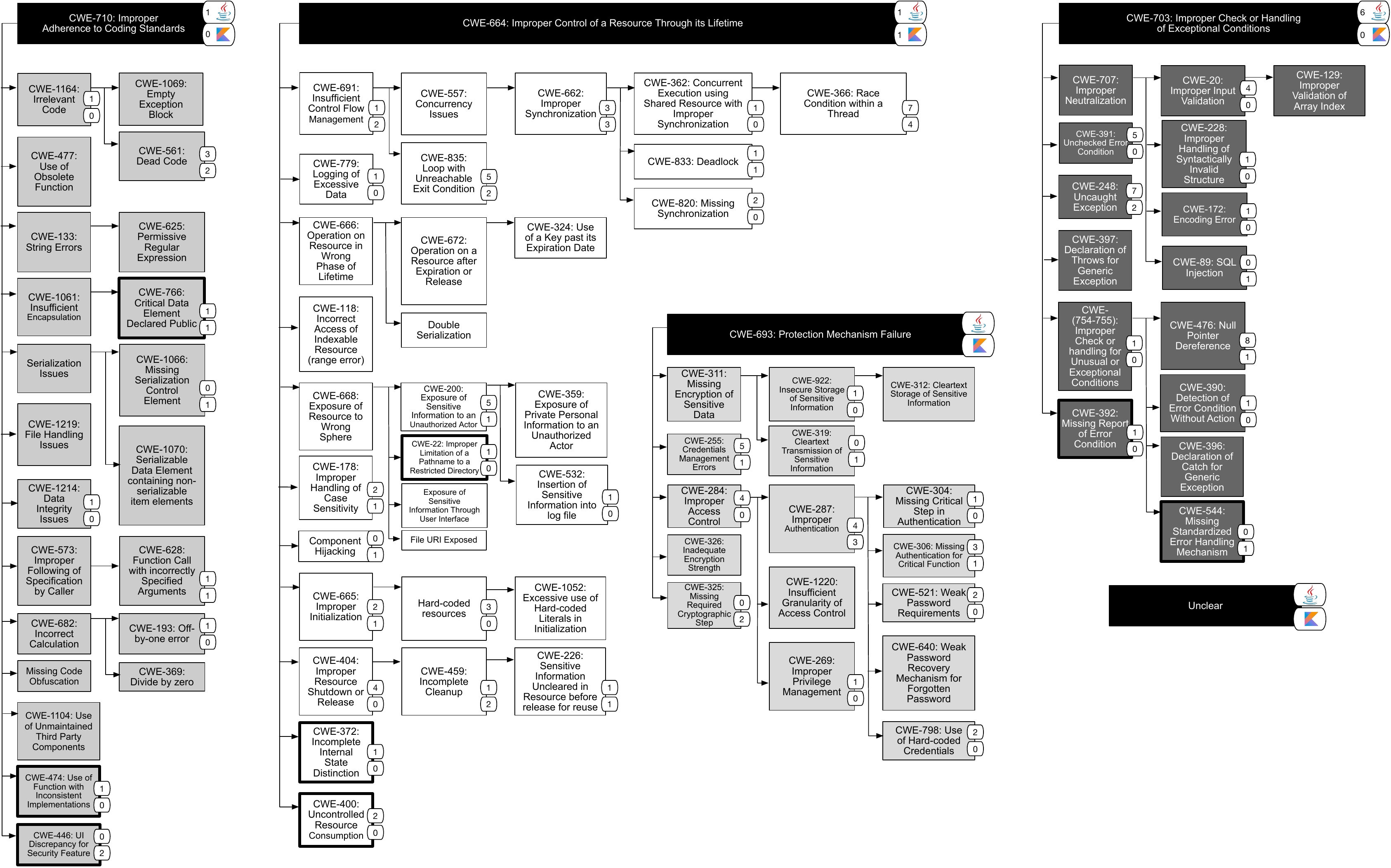}
		\caption{Validation taxonomy of types of security weaknesses found in Java and Kotlin Android apps.}
		\label{fig:taxonomyValidation}
	\end{center}
\end{figure*}

Among the 250 commits analyzed by the two Master students (see \secref{sec:design} for details), 73 were classified as false positives for Java and 24 for Kotlin. This left us with 153 valid instances that have been used for the construction of the validation taxonomy (See \figref{fig:taxonomyValidation}). Looking at it, it can be seen that 85\% of the identified categories are already covered in our taxonomy and only 8 new categories were identified (\ie \emph{CWE-22: Improper Limitation of a Pathname to a Restricted Directory}, \emph{CWE-372: Incomplete Internal State Distinction}, \emph{CWE-392: Missing Report of Error Condition}, \emph{CWE-400: Uncontrolled Resource Consumption}, \emph{CWE-446: UI Discrepancy for Security Feature}, \emph{CWE-474: Use of Function with Inconsistent Implementations}, \emph{CWE-544: Missing Standardized Error Handling Mechanism}, and \emph{CWE-766: Critical Data Element Declared Public}). Also, all these categories are child of one of our root categories. This indicates a good generalizability of our taxonomy. Additionally, although the proportion of Kotlin artifacts is considerably lower than the amount of Java ones, it is worth noting that in the two taxonomies the distribution of types of security weaknesses across Java and Kotlin is similar.

\begin{table*}[tb]
\centering
\caption{Security weaknesses mentioned by developers: Available tools\vspace{-0.2cm}}
\label{tab:toolsSurveyess}
\rowcolors{2}{gray!15}{white}
\begin{tabular}{p{8.2cm}|p{9.2cm}}

\toprule

\textbf{Security weaknesses} & \textbf{Tools} \\ \midrule

CWE-20: Improper Input Validation 
& DifFuzz \cite{DifFuzz}, DroidFuzzer \cite{DroidFuzzer}, EvoTaint \cite{EVOTAINT}, Flowdroid \cite{arzt2014flowdroid}, Huang \etal \cite{Huang2019fuzzing}, IVDroid \cite{IVDROID}, Monkey \cite{monkey}\\

CWE-89: SQL Injection & OPIA \cite{bello2019opia}, Kul \etal \cite{KUL}\\ 
CWE-200: Exposure of Sensitive Information to an Unauthorized Actor & AppFence \cite{AppFence}, AppIntent \cite{AppIntent}, AutoPatchDroid \cite{AutoPatchDroid}, Blackdroid \cite{Blackdroid}, CoChecker \cite{CoChecker}, ComDroid \cite{ComDroid}, ContentScope \cite{ContentScope}, Covert \cite{bagheri2015covert}, CredMiner \cite{CredMiner},Flowdroid \cite{arzt2014flowdroid}, IccTA \cite{li2015iccta}, Kul \etal \cite{KUL}, Matsumoto2013 \etal \cite{Matsumoto2013}, MITHYS \cite{MITHYS}, M-Perm \cite{MPERM}, OAUTHLINT \cite{OAUTHLINT}, Onwuzurike \etal \cite{Onwuzurike2015}
\\ 
CWE-269: Improper Privilege Management & AppProfiler \cite{AppProfiler}, AppGuard \cite{AppGuard}, AutoPatchDroid \cite{AutoPatchDroid}, AWiDe \cite{AWiDe}, Bartsch \etal \cite{Bartsch2013}, CoChecker \cite{CoChecker}, Covert \cite{bagheri2015covert}, DroidChecker \cite{DroidChecker}, Droidtector \cite{Droidtector}, Lintent \cite{Lintent}, M-Perm \cite{MPERM}, PaddyFrog \cite{PaddyFrog} 
\\ 
CWE-284: Improper Access Control & ContentScope \cite{ContentScope}\\ 
CWE-311: Missing Encryption of Sensitive Data & DroidSearch \cite{DroidSearch}, OPIA \cite{bello2019opia}\\
CWE-325: Missing Required Cryptographic Step & CrypLint \cite{egele2013empirical}\\ 
\makecell[tl]{CWE-312: Cleartext Storage of Sensitive Information \\ CWE-922: Insecure Storage of Sensitive Information} & Blackdroid \cite{Blackdroid}, CredMiner \cite{CredMiner}, Flowdroid \cite{arzt2014flowdroid}\\ 
\makecell[tl]{CWE-359: Exposure of Private Personal Information \\ to an Unauthorized Actor \\ CWE-798: Use of Hard-coded Credentials} & Flowdroid \cite{arzt2014flowdroid}, Kul \etal \cite{KUL}, M-Perm \cite{MPERM}, CredMiner \cite{CredMiner}\\ 
Component Hijacking & ActivityHijacker \cite{ActivityHijacker}, AppSealer \cite{AppSealer}, CHEX \cite{CHEX}, ComDroid~\cite{ComDroid}, Ren \etal \cite{ren2015hijacking}, You \etal \cite{you2016reference}\\ 

\bottomrule

\end{tabular}
\vspace{-0.3cm}
\end{table*}

%% file: threats.tex
\section{Threats to Validity} \label{sec:threats}
\vspace{-0.2cm}

\textbf{Construct validity.} We identified through manual analysis the types of security weaknesses fixed by developers. To mitigate subjectivity bias, two authors have been assigned to each commit and, in case of conflict, the commit was assigned to a third evaluator. 

Also, when the type of security flaw being fixed was not clear, we assigned the ``unclear'' tag rather than best-guessing the classification. Despite this mitigation strategies, imprecisions are still possible.

Concerning the survey, we tried to not bias the participants' answers especially in the context of questions asking for the most common/dangerous security weaknesses they faced in their apps. For this reason, we did not provide a multiple choice answer but we used an open answer.

\textbf{Internal validity.} In the survey, we collected information about the background of the participants, and excluded developers having no experience with native Android apps. For the manual study, we acknowledge that we only analyzed one specific source of information (\ie security weakness-fixing commits) and this may have an impact on the derived taxonomy. Similarly, we only included in the manual analysis commits that impacted a single file, to make sure that the ``security weakness'' mentioned in the commit message was located in that file. Again, this could have affected the resulting taxonomy.

\textbf{External validity.} We manually analyzed a total of \totalLabeled security weakness-fixing commits coming from \totalApps apps. However, due to the removal of false positives and ``unclear'' instances, our taxonomy is based on 386 actual instances. Also, we asked two Master students to analyze an additional set of 250 instances to test the generalizability of our taxonomy. Analyzing additional instances and other orthogonal sources of information (\eg \url{cvedetails.com}) could complement our taxonomy. As for the survey, we collected a total of \participants complete answers. While this number is limited, it is in line with many previously published survey studies in software engineering (see \eg \cite{DagenaisOBRV10,CanforaPOP12,Romano:tse2020}).

%% file: related.tex

\vspace{-0.2cm}
\section{Related Work} \label{sec:related}


Several techniques have been proposed to detect, and in some cases fix, vulnerabilities in mobile apps (\eg~\cite{arzt2014flowdroid,li2015iccta,sadeghi2017taxonomy,lee2017sealant,singleton2019firebugs,you2016reference, bello2019opia}). 
We focus on studies investigating security-related aspects in \emph{Android apps}, since these are the most related to our work. \textRevision{\tabref{tab:relatedWorks} provides an overview of the discussed studies, reporting for each \emph{reference}, the \emph{year of publication}, a \emph{brief summary} of its contribution, the \emph{size} of the dataset including the number of analyzed apps (\#a) or commits (\#c) since our paper reports this information, along with the number of \emph{security weaknesses types} and \emph{categories} that have been outlined.
}

\begin{table*}[tb]
\vspace{0.4cm}
\centering
\caption{\textRevision{Empirical studies on security weaknesses in Android apps}}
\label{tab:relatedWorks}
\rowcolors{2}{gray!15}{white}
\begin{tabular}{l|l|l|c|c|c}

\midrule

\textbf{Ref.} & \textbf{Year} & \textbf{Brief summary} & \textbf{\textRevision{Size}} & \textbf{\textRevision{Types}} & \textbf{\textRevision{Categories}} \\ \midrule

\multicolumn{1}{c|}{\cite{felt2011android}} & 2011 & Detection of overprivileges in Android apps & \texttt{\#a:} 940 & 10 & 1 \\ 
\multicolumn{1}{c|}{\cite{enck2011study}} & 2011 & Identification of vulnerabilities' root causes & \texttt{\#a:} 1,100  & 8  & 1\\ 
\multicolumn{1}{c|}{\cite{egele2013empirical}} & 2013 & Cryptographic misuse in Android apps & \texttt{\#a:} 11k+  & 6 & 1 \\ 
\multicolumn{1}{c|}{\cite{zuo2015automatically}} & 2015 & Detection of SSL error-handling vulnerabilities &  \texttt{\#a:} 13,820 & 1 & 1 \\
\multicolumn{1}{c|}{\cite{bagheri2015covert}} & 2015 & Analysis of inter-app security vulnerabilities &  \texttt{\#a:} 500 & 2 & 1 \\ 
\multicolumn{1}{c|}{\cite{ahmad2016inter}} & 2016 & Developers challenges for inter-app communication  & \texttt{\#a:} 52k & 3 & 1 \\ 
\multicolumn{1}{c|}{\cite{weir2020needs}} & 2020 & Survey on developer practices for app security & \texttt{\#a:} 454 & 3 & 1 \\ 
\multicolumn{1}{c|}{\cite{gao2021understanding}} & \textRevision{2021} & \textRevision{Temporal evolution of vulnerabilities in Android apps} & \texttt{\#a:} 465,037 & 10  & 4 \\ 
\hline
\multicolumn{2}{c|}{This paper} & \textRevision{Taxonomy of Security Weaknesses} &\texttt{\#a:}8,157  \texttt{\#c:}4,781 & 80 & 5 \\ 


%
%
%





\bottomrule

\end{tabular}
\end{table*}

Felt \etal~\cite{felt2011android} identified over-privileges in the permissions (\eg bluetooth, read contacts) of one-third of the 940 Android apps they analyzed. \textRevision{10 most common unnecessary permissions are identified, and the percentage of overprivileged applications varies from 5\% to 16\%.} The authors point out that this is mainly due to developers not interpreting correctly the API documentation. The results of our work, and especially of our survey, support the relevance of permissions for the vulnerabilities affecting Android apps.

Enck \etal~\cite{enck2011study} investigated the root causes of vulnerabilities in 1,100 free Android apps. The authors find misuse of sensitive information (\ie phone identifiers and geographic location) among the root causes. \textRevision{Android-specific vulnerabilities relate all to the sensitivity of data, and 8 different types are identified, \eg leaking information to logs, unprotected broadcast receivers, etc.} The security of Android APIs was also considered insufficient, but no vulnerability was found able to maliciously control the apps. 

The mishandling of sensitive information is also a prevalent aspect in our taxonomy.

Egele \etal~\cite{egele2013empirical} used a static analysis technique to capture cryptographic misuses in 11k+ apps. They showed that 88\% of the analyzed apps do not correctly use cryptographic APIs. This is mainly due to the lack of inter-procedural analysis that correlates multiple functionalities (\eg encryption and decryption) within a method instantiation. \textRevision{Focus here is on cryptography, and 6 different types of violations (\eg constant encryption keys) have been highlighted.} Instances of issues related to cryptography are found both in Java and Kotlin in our taxonomy.

Sufatrio \etal~\cite{tan2015securing} presented a secondary study reviewing the literature about existing security solutions for Android apps. The taxonomy is relevant for five deployment stages, \ie development, availability on markets, installation on a device, execution, and security settings modification. \textRevision{It surveys existing work, it does not rely on a specific dataset of analyzed apps/commits, but it elaborates on the literature to derive a taxonomy including 5 categories and 18 types of security vulnerabilities that should be prevented.}

Zou \etal \cite{zuo2015automatically} exploited static and dynamic analysis to detect apps opening {\tt https} web pages with illegal certificates. \textRevision{It targets a specific category of vulnerabilities, \ie the privacy of the communications. The developed framework detects a specific type of violation, \ie ignoring the illegal certificate error and proceeding with the sending of sensitive information over an insecure communication channel.} 
Bagheri \etal ~\cite{bagheri2015covert} analyzed inter-app and inter-component security vulnerabilities in 500 apps. Specifically, a formal model expressing security properties of apps/components is extracted and a model checker verifies the safety of simultaneously running two apps that may interact while holding certain permissions. \textRevision{This research focuses on identifying a specific category of vulnerability, \ie privilege escalation --an application with less permissions can be not restricted to access components of a more privileged application--. Two types of detection strategies are adopted: (i) entities that can be inferred from a method; (ii) vulnerable paths of communication between entities.} Also Ahmad \etal \cite{ahmad2016inter} analyzed inter-app communication (IAC) in 52k apps\textRevision{, where the focus is on  different types of actors involved  in IAC (Library, Caller, and Callee), which  are recognized as types of entities potentially vulnerable. Overall, these works \cite{zuo2015automatically,bagheri2015covert,ahmad2016inter} focus on a specific category of security vulnerabilities that, also due to the nature of our investigation (intentionally meant to be more general)}, we did not identify in our study.
%

%

Android devices and the operating system have been also investigated. Meng \etal~\cite{meng2018survey} presented a taxonomy of \textRevision{63} device exploits (\ie vulnerabilities leading to privilege escalation) \textRevision{grouped in 3 main categories that are related to perspectives: societal, practical, and technical. It is shown} 
that the diffusion of exploits is decreasing due to Android systems and Linux kernels strengthening their security mechanisms. Our study does not limit its focus to exploits, but looks at security weaknesses from a more general perspective.

Jimenez \etal~\cite{jimenez2016profiling} presented a taxonomy of 43 issues related to Android OS vulnerabilities by leveraging the CVE-NVD (Common Vulnerability Exposures - National Vulnerability Database) database \textRevision{whose size is left unspecified}. The authors found that Android vulnerabilities \textRevision{related to the code mainly belong to 9 categories (\eg resource management, handling data, \etc). They} are mainly located in components dealing with browsing, cryptography, access control or networking. Also the fixing of vulnerabilities is investigated looking at the distribution of code changes, and most of them related to the additions of condition(s), authorization, functions, \etc Mazuera-Rozo \etal~\cite{mazuera2019android} also performed empirical studies on the Android OS to categorize the types of the vulnerabilities (\eg denial of service, improper authorization), their evolution overtime and their survivability. \textRevision{Security weaknesses are grouped in 14 categories where 154 types (\eg credentials management, improper authorization, transmission of sensitive information, \etc) have been identified.} Besides, vulnerability patches (\eg check for exceptional conditions, proper handling of certificates, appropriate initialization values for variables) are analyzed to investigate the most used fixes. Our work, while related, focuses on security weaknesses affecting Android apps rather than the Android OS.

%

Weir \etal~\cite{weir2020needs} conducted a survey on the effect of requirements and developer practices on apps' security. For app development, security is perceived relevant by the participants, even if assurance techniques are poorly used. \textRevision{The survey refers to a set of 454 apps, and 335 developers were using tools suitable to check the following 3 types of weaknesses: SSL security, cryptographic API misuse, and privacy leaks. As result, a portion of participants have been classified as security specialists and they advocated the usage of cryptography to enforce security.}

\textRevision{Gao \etal~\cite{gao2021understanding} investigated the temporal evolution of vulnerabilities in Android apps. Vulnerable code is detected in terms of which locations (\eg library code) are more common than others, the types of code change (\eg the addition of new files) that may entail security-related issues, and also if there is a correlation with malwares. The list of considered vulnerabilities is constituted of 4 categories (\ie security features, permissions, injection flaws and data/communication handling) and 10 types, each associated to a detection tool providing evidence of the corresponding vulnerability.} 

To the best of our knowledge, our work represents the first and most comprehensive taxonomy of security weaknesses in Android apps, including both Java and Kotlin app-related code. Besides, our taxonomy is the result of a two-phase study, involving the inspection of software-related artifacts (\ie security weakness-fixing commits) and a survey with software developers. The derived taxonomy is more comprehensive and extensive, covering 18 of the 20 issues analyzed in previous papers by Enck \etal\cite{enck2011study}, Egele \etal\cite{egele2013empirical}, Zuo \etal\cite{zuo2015automatically}, Bagheri \etal\cite{bagheri2015covert}, Jimenez \etal\cite{jimenez2016profiling}, Weir \etal\cite{weir2020needs}\textRevision{, and Gao \etal~\cite{gao2021understanding}}. Finally, we focus on both Java and Kotlin code \textRevision{recently suggested in~\cite{coppola2019migrationkotlin}}, while only Java-related security weaknesses are analyzed in previously mentioned works.

%% file: conclusion.tex

\vspace{-0.4cm}
\section{Conclusions} \label{sec:conclusion}
\vspace{-0.15cm}

We presented the first available taxonomy of security weaknesses in Android apps that covers both Java- and Kotlin-related code. Our taxonomy features 80 types of software security weaknesses, and it is the result of both a mining-based study in which we manually inspected \totalLabeled  commits fixing security weaknesses (that contributed 74 types of security weaknesses), and a survey performed with \participants developers (contributing six additional types). \textRevision{Our results discussion resulted in the identification of several lessons learned for both practitioners (see the \faCodeFork~icon in Section 3) and researchers (\faFlask~icon).}

Our future work will be mostly driven by the findings discussed in \secref{sec:results}. In particular, we plan to focus on the definition of techniques able to detect (and possibly automatically fix) security weaknesses that are (i) not currently supported by existing detection tools, (ii) frequently spread in real Android apps, and (iii) relevant for software developers. Besides, we are interested to investigate the portability of methodologies and tools detecting Java-based weaknesses in Kotlin-based code, to understand which changes are needed to enable interoperability between the two languages. Our study provides the foundations for such a research agenda. 


%% file: ack.tex

\vspace{-0.3cm}
\section*{Acknowledgment} \label{section:ack}
\vspace{-0.1cm}
Mazuera-Rozo and Bavota gratefully acknowledge the financial support of the Swiss National Science Foundation for the CCQR project (SNF Project No. 175513). Escobar-Vel\'asquez and Linares-V\'asquez were partially funded by a Google Latin America Research Award 2018-2021. Escobar-Velásquez was supported by a ESKAS scholarship, No. 2020.0820. Trubiani was partially supported by the MIUR PRIN project 2017TWRCNB SEDUCE.